\def\cm{{\rm\thinspace cm}}

\def\erg{{\rm\thinspace erg}}

\def\keV{{\rm\thinspace keV}}

\def\Msun{\hbox{$\rm\thinspace M_{\odot}$}}

\def\s{{\rm\thinspace s}}

\def\Hz{{\rm\thinspace Hz}}

\def\ergcmps{\hbox{$\erg\cm\ps\,$}}

\def\pcmsq{\hbox{$\cm^{-2}\,$}}

\def\ps{\hbox{$\s^{-1}\,$}}
\def\psqcm{\hbox{$\cm^{-2}\,$}}

\documentclass[usegraphicx]{mn2e}

\usepackage{fixltx2e}
\usepackage{mathptmx}

\include{defn}
\newcommand{\oneh}{{1H\,0707-495}}
\newcommand{\xrt}{{\it Swift-XRT}}

\begin{document}

\title[1H\,0707-495 in 2011]{1H\,0707-495 in 2011: An X-ray source within
  a gravitational radius of the event horizon} \author[Fabian et al]
{\parbox[]{6.5in}{{A.C. Fabian$^1\thanks{E-mail: acf@ast.cam.ac.uk}$,
      A. Zoghbi$^1$, D. Wilkins$^1$, T. Dwelly$^2$, P. Uttley$^2$,
      N. Schartel$^3$, G.Miniutti$^4$, L. Gallo$^5$, D. Grupe$^6$, 
S. Komossa$^7$ and M. Santos-Lle\'o$^3$
    }\\
    \footnotesize
    $^1$ Institute of Astronomy, Madingley Road, Cambridge CB3 0HA\\
$^2$ School of Physics and Astronomy, University of Southampton,
Highfield, Southampton SO17 1BJ\\
$^3$ XMM-Newton Science Operations Centre, ESA, Villafranca del
Castillo, Apartado 78, E-28691 Villanueva de la Ca{\~n}ada, Madrid, Spain\\
$^4$ Centro de Astrobiologia (CSIC-INTA), Dep. de Astrofisica; LAEFF, PO Box 78, E-28691, Villanueva de la Ca{\~n}ada, Madrid, Spain\\
$^5$ Department of Astronomy and Physics, Saint
Mary’s University, Halifax, NS B3H 3C3, Canada\\
$^6$ Department of Astronomy and Astrophysics, Pennsylvania
State University, 525 Davey Lab, University Park,
PA 16802, USA\\
$^7$ Max-Planck-Institut f\"ur extraterrestrische Physik, Giessenbachstr., D-85748 Garching, Germany\\
  } }

\maketitle
  
\begin{abstract}
  The Narrow Line Seyfert 1 Galaxy 1H\,0707-495 went in to a low state
  from 2010 December to 2011 February, discovered by a monitoring
  campaign using the X-Ray Telescope on the Swift satellite. We
  triggered a 100~ks XMM-Newton observation of the source in 2011
  January, revealing the source to have dropped by a factor of ten in
  the soft band, below 1~keV, and a factor of 2 at 5~keV, compared
  with a long observation in 2008. The sharp spectral drop in the
  source usually seen around 7~keV now extends to lower energies,
  below 6~keV in our frame. The 2011 spectrum is well fit by a
  relativistically-blurred reflection spectrum similar to that which
  fits the 2008 data, except that the emission is now concentrated
  solely to the central part of the accretion disc. The irradiating
  source must lie within 1 gravitational radius of the event horizon
  of the black hole, which spins rapidly. Alternative models are
  briefly considered but none has any simple physical interpretation.
\end{abstract}


\section{Introduction}

The Narrow-Line Seyfert 1 Galaxy 1H\,0707-495 is bright in soft X-rays
where it shows the steep spectrum and rapid variability typical of its
class. XMM-Newton observations revealed a sharp drop in its spectrum
around 7~keV, which can be interpreted either as an iron absorption
feature or as the blue wing of a large broad emission line (Boller et
al 2002; Fabian et al 2002). Further observations showed the drop to
shift to a higher energy (Gallo et al 2004), possibly due to
ionization changes. A 500~ks XMM observation of 1H\,0707-495 in 2008
provided clear support for the emission interpretation. A feature at
1~keV enables the spectrum to be decomposed into a power-law continuum
with broad Fe-K and Fe-L emission lines (Fabian et al 2009). These are
expected from irradiation of an iron-rich accretion disc around a
black hole and are well fitted by a relativistically-blurred
reflection model (Fabian et al 2009; Zoghbi et al 2010).

High frequency variations in the power-law continuum are followed
about 30~s later by variations in the reflection component (Fabian et
al 2009; Zoghbi et al 2010). Such reverberation and its spectrum
(Zoghbi et al 2011) are fully consistent with, and expected from,
reflection. The data indicate a rapidly spinning black hole of mass
$M\sim 5\times 10^6\Msun$ and spin $a>0.97$. The power-law continuum
originates from a corona close ($<10$~ r$_{\rm g}=10 GM/c^2$) to the
black hole. 

We report here on new XMM observations made in 2011 January in
response to a prolonged drop in X-ray flux from 1H\,0707-495 seen in a
monitoring campaign of the source made with the X-Ray Telescope on {\em
  Swift}. The spectrum shows the source at an unprecedented low level
in the soft X-ray band. A reflection spectrum provides a good fit, and
requires that the power-law continuum source has moved to within a radius
2\,r$_{\rm g}$ of the black hole. 

Alternative interpretations are briefly discussed. Deep XMM-RGS
observations in 2008 show no evidence for a line-of-sight ionized wind
or warm absorber in this object (Blustin \& Fabian 2009) and we do not
pursue such models further here.

Other AGN have been seen in a low state with strong reflection
components (e.g. 1H\,0419-577, Fabian et al 2005; NGC\,4051, Ponti et al 2006; 
PG\,1543+489, Vignali et al 2008; Mrk\,335, Grupe et al 2008;
PG\,1535+547. Ballo et al 2008; PG\,2112+059, Schartel et al
2010). The spectrum of 1H\,0707-495 presented here has the most
extreme relativistic parameters yet seen.
 
\section{SWIFT monitoring campaign}
\begin{figure}
  \centering
  \includegraphics[width=1.1\columnwidth,angle=0]{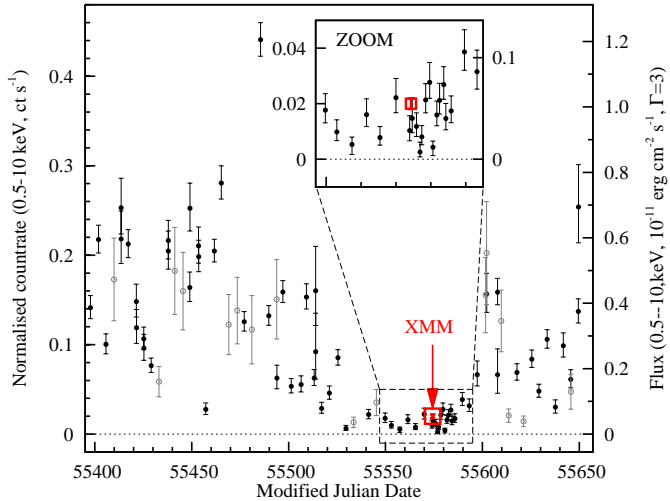}
  \caption{The \xrt\ lightcurve for \oneh\ showing the pronounced dip in
  X-ray brightness.  Each point shows the normalised 0.5--10\,keV
  countrate measured during a single \xrt\ visit. An approximate flux
  scale is indicated, derived assuming a powerlaw spectrum with slope
  $\Gamma = 3$, and corrected for the Galactic column.  A systematic
  uncertainty of 25 per cent has been added in quadrature to measurements
  where the target position landed on or near bad CCD columns (shown
  with open grey symbols).  The epoch of the {\em XMM-Newton}
  observation is indicated with a vertical arrow. 
 }
\end{figure}
\begin{figure}
  \centering
  \includegraphics[width=0.7\columnwidth,angle=-90]{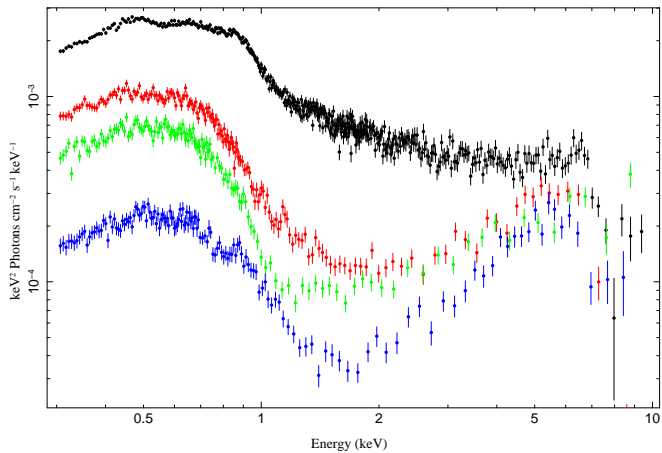}
  \caption{XMM pn spectra of 1H\,0707-495 in 2008 (black, upper),  2000
    (red), 2007 (green) and 2011 (blue, lower).}
\end{figure}

During the period 2010 March to 2011 March we monitored the X-ray and
UV emission from \oneh\ using the {\em Swift} satellite, in order to
improve measurements of the low frequency power spectrum of the
source.  A $\sim$1\,ks observation was obtained every $\sim$4 days. In
some cases, due to observational constraints, the 1\,ks exposure time
was split into two or more short sub-exposures (``visits'').  We used
an automated script to download the \xrt\ data from the HEASARC quick
look website\footnote{http://swift.gsfc.nasa.gov/cgi-bin/sdc/ql}
shortly after they were obtained. The \xrt\ data were then processed
and reduced through our automated pipeline, which combines a mixture
of standard HEASOFT
tools\footnote{http://heasarc.gsfc.nasa.gov/docs/software/lheasoft/}
and some of our own routines. We outline the pertinent steps of our
pipeline below.  Firstly the raw \xrt\ data is reprocessed with the
{\sc xrtpipeline} tool (version 0.12.6) to produce a calibrated and
cleaned events file.  A zeroth order measure of the countrate is
obtained by constructing a lightcurve (using the {\sc xselect} tool)
from the counts detected within a 30~arcsec radius aperture centered
on the nominal position of \oneh.  Following the method of Evans et
al. (2007) we then maximize the signal to noise ratio by using an
adaptively sized aperture which is chosen to give the highest quality
measurements of the the source countrate. The radius of the optimally
sized circular aperture is determined from the zeroth order
brightness, and for \oneh\ ranged between 17 and 45~arcsec. The
background is estimated by measuring the countrate within an annular
region surrounding the target aperture, and is subtracted taking
account of the relative areas of the source and background regions.
The effective area of \xrt\ is strongly sensitive to the location of
the source within the image plane and to the specific extraction
aperture used. The following procedure is used to normalize the raw
countrate.  Firstly the {\sc xrtmkarf} tool is used to calculate an
ancillary response function (ARF) file for each visit, taking account
of the exposure map at the location of the source and the aperture
size used.  Using this ARF together with the standard XRT response
matrix, and adopting a spectral model of a $\Gamma = 3$ powerlaw
absorbed by the Galactic column, we generate a fake visit spectrum
using {\sc XSPEC}. We then calculate the appropriate normalisation
factor by dividing the countrate of the faked visit spectrum by the
countrate expected from a faked spectrum generated using the nominal
instrumental ARF. Using the faked spectrum associated with the nominal
ARF we calculate an approximate normalised countrate to flux
conversion factor of
0.366\,ct\,s$^{-1}$\,$(10^{-11}$\,erg\,cm$^{-2}$\,s$^{-1}$)$^{-1}$ for
the 0.5--10\,keV band.  Uncertainties on countrates are estimated
assuming Poisson statistics, following the method of Gehrels
(1986). Note that during a number of visits, the target position
landed directly on, or very close to, one of several bad columns on
the XRT detector. In these cases the normalisation factor is somewhat
uncertain, and so we have added in quadrature a systematic uncertainty
of 25 per cent.

The progress of the {\em Swift} monitoring was checked regularly, and
we noticed early in 2011 January  (around MJD 55570) that \oneh\ had
become exceptionally faint in the X-ray band.  A {\em Swift} ToO
request was then initiated with daily monitoring of \oneh\ which
confirmed the prolonged dip in the X-ray brightness.  A section of the
0.5--10\,keV \xrt\ lightcurve, bracketing the pronounced dip in X-ray
flux, is presented in Fig.~1.

The Optical Monitors on both {\em Swift} and XMM-Newton both record
the UV flux of the source. No marked change was seen during the
observations (Cameron et al, in preparation). 

\section{XMM-Newton data from 2011}

1H0707-495 was observed with XMM-Newton on 2011 January 12--13
(Obs. ID 0554710801) in large window mode for a total exposure of 100
ks. The observational data files (odf) were reduced using XMM-Newton
science software {\sc sas v11.0.0}. There were some strong background
flares during the observation, which were excluded in the spectral
analysis leaving a total net exposure of 65 ks. Standard filtering and
event selection were applied. The source spectrum was extracted from a
circular region of radius 35 arcsec centred on the source, and
background from the the surrounding regions in the same chip. The
spectra were then grouped using {\sc grppha} so that each bin has a
minimum of 50 counts. The response files were generated using {\sc
  rmfgen} and {\sc arfgen}. We show only pn data here, for ready
comparison to the 2008 analysis and since the pn has better
sensitivity in the Fe-K band. The MOS data show good agreement with
the pn results.

\begin{figure}
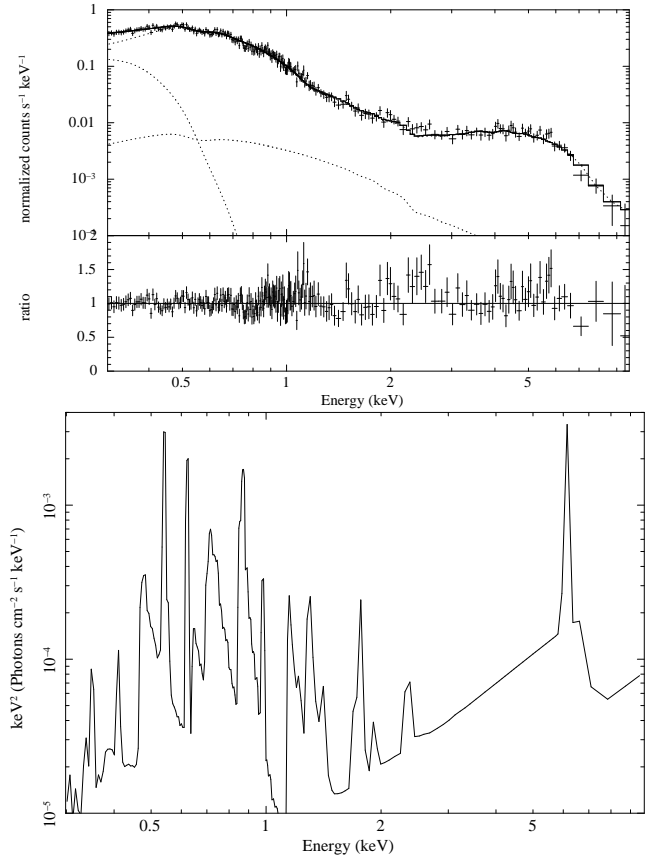

  \centering
  \includegraphics[width=0.65\columnwidth,angle=-90]{best1refladd.ps}
  \includegraphics[width=0.7\columnwidth,angle=-90]{wareflpo.ps}
  \caption{Top: XMM pn spectrum from 2011 with best-fitting model
    consisting of a blurred reflection component, power-law continuum
    (the best-fitting amplitude is shown but the data are consistent
    with zero) and soft blackbody (see Table 1 for details). Lower
    panel: unblurred reflection model.}
\end{figure}
\begin{figure}
  \centering
  \includegraphics[width=1\columnwidth,angle=0]{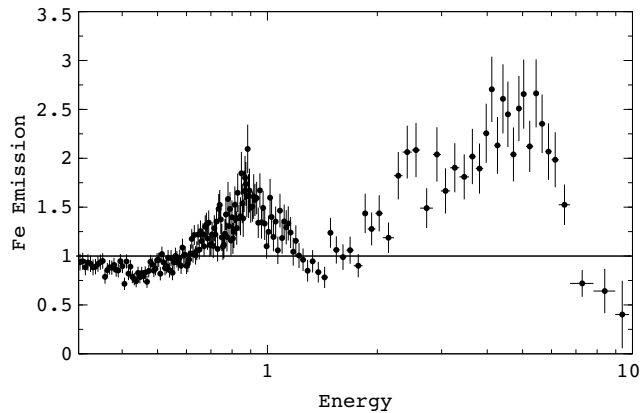}
  \caption{Contribution of iron emission to the single reflector
    model. Note the strong contributions at both Fe-K and Fe-L. }
\end{figure}
\begin{figure}
  \centering
  \includegraphics[width=0.7\columnwidth,angle=-90]{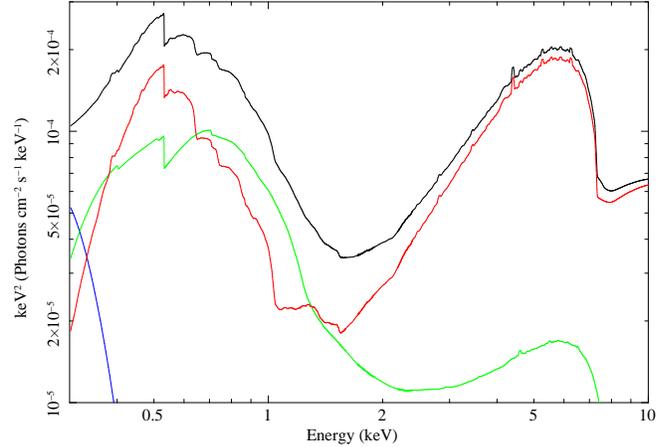}
  \caption{Double reflector model. The high ionization component is in
    green and the low ionization component in red. The reflectors
    differ only in ionization parameter and normalisation.}
\end{figure}
\begin{figure}
  \centering
  \includegraphics[width=1\columnwidth,angle=0]{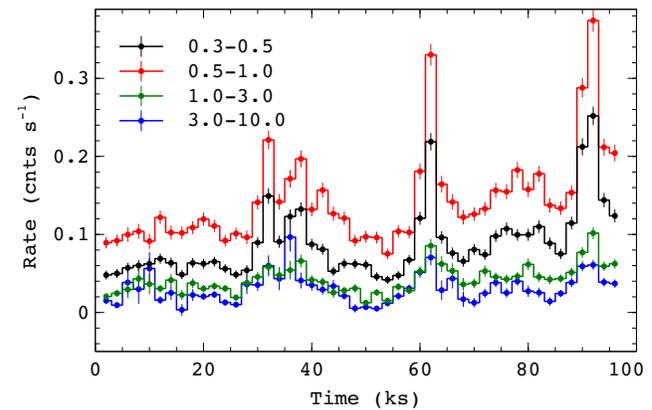}
  \caption{Multicolour X-ray pn lightcurves for the XMM 2011 observation.}
\end{figure}
\begin{figure}
  \centering
  \includegraphics[width=1\columnwidth,angle=0]{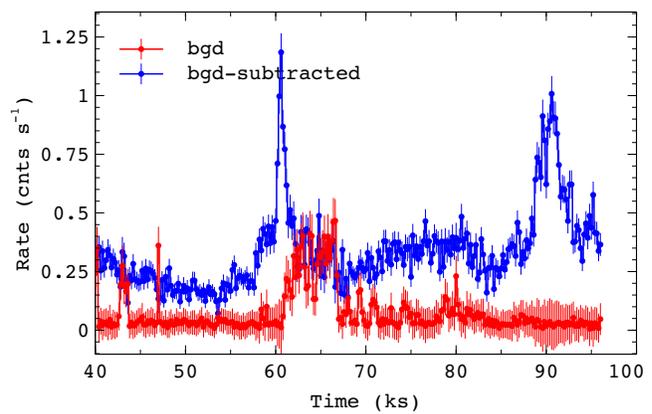}
  \caption{Lightcurve of whole band from the pn detector, with 200\,s
    binning, over the interval showing the highest flares.The
    background count rate of a larger spatial region, scaled to the
    extraction region used for 1\,H0707-495, is shown in red.}
\end{figure}

\begin{figure}
  \centering
  \includegraphics[width=1\columnwidth,angle=0]{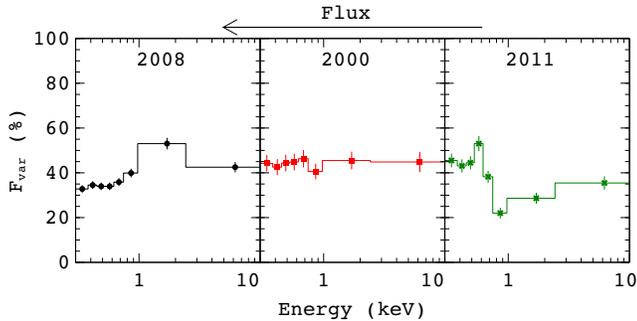}
  \caption{Fractional variability as a function of energy for 3
    different observations of 1H\,0707-495. }
\end{figure}

\begin{figure}
  \centering
  \includegraphics[width=1\columnwidth,angle=0]{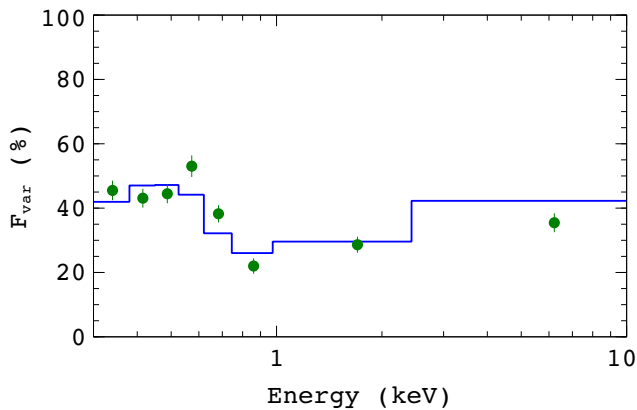}
  \caption{Fractional variability from 2011 observation plotted with
    model in which the variability is due to ionization changes in the
    lowly ionized reflection component.}
\end{figure}
\subsection{Model Fitting}

\begin{table}
  \caption{Values of  model parameters used in emisivity profile
    determination.The model used is {\tt phabs*zphabs*(blackbody+
      kdblur*atable\{reflionx.mod\})}. The two absorption components
    have a fixed Galactic value and a variable intrinsic one,
    respectively. Representative values from fits to the total 2008 data (Zoghbi
    et al 2011) are listed
    in the last column for comparison with the new results. The source
    varied considerably during the 500~ks observation.}
\centering
\begin{tabular}{llll}
  	\hline
   	\textbf{Component} & \textbf{Parameter} & \textbf{Value} & 2008 \\
	\hline
        Absorption & Galactic $N_{\rm H}\pcmsq$ & $4.3 \times 10^{20}$ & \\
        & Intrinsic $N_{\rm H}$ & $7.1 \times 10^{20}$ & \\
        \hline
	powerlaw  & Photon index, $\Gamma$ & $2.71^{+0.1}_{-0.07}$ & 3.2
        \\
        & Norm &  $<1.3\times 10^{-5}$ & $1.1\times 10^{-3}$ \\
	\hline
	kdblur  & Inclination, $i$\,deg & $69.3^{+1.2}_{-2.2}$ & 58.5 \\
	& $R_{\rm in}$\,r$_{\rm g}$ & $<1.3$ & 1.23 \\
        & Index, $q$ & $>8.6$ & 6.6 \\
        \hline
        blackbody & temperature, $kT$\,keV & $0.030^{+0.01}_{-0.007}$
        & 0.037  \\
        & Norm & $5.7^{+25}_{-4.9}\times 10^{-4}$ & $2.1\times 10^{-3}$\\
        \hline
	reflionx  	& Iron abundance / solar & $>7.2$ & $>7$ \\
	& Ionisation parameter, $\xi$ & $12.7^{+2.8}_{-1.7}$ & 57 \\
        & $\ergcmps$ & & \\
        & Norm & $1.5^{+0.9}_{-0.6}\times 10^{-5}$ & $5.1\times 10^{-5}$\\ 
	& Redshift, $z$ & $4.06\times 10^{-2}$ & \\
	\hline
        $\chi^2/{\rm dof}$ & $317/340$ & \\
        \hline
\end{tabular}

\label{par.tab}

\end{table}

The XMM-Newton pn spectrum is shown in comparison to that of 2008 in
Fig.~2. It is clear that there is a large, order of magnitude, drop in
the soft flux below 1~keV, but only a factor of about 2 drop at 5
keV. In an $EF(E)$ spectral sense the hard flux peaks between 5 and
6~keV. Other XMM observations of the source give spectra lying between
those of 2008 and 2011 (Fig.~2), but at least a factor of 3 above that
of 2011 in the soft band.


We fit the 0.3--10~keV spectrum with a model similar to that used by
Fabian et al (2009) and Zoghbi et al (2010), consisting of a
relativistically-blurred reflection component (convolution model {\sc
  kdblur} acting on ionized reflection model {\sc reflionx} of Ross \&
Fabian 2005)) and a low temperature blackbody. Absorption by a
Galactic column density of $4.3\times 10^{20}\psqcm$ (Kalberla et al
2005) is applied and intrinsic cold absorption is allowed in the
fitting process. The blackbody emission in only important below
0.5~keV. There is no evidence of the direct power-law continuum. The
parameters of the model are given in Table 1 and the fit is shown in
Fig.~3. (The MOS data give consistent results.)  Overall, the
parameters are similar to those fitting the 2008 data, except that the
photon index $\Gamma$ is flatter and the inclination is $\sim10$~ deg
larger. The strong contribution of iron emission to the spectrum is
demonstrated in Fig.~4, which was obtained by dropping the iron
abundance in the model to its lowest value, 0.1, plotting the ratio of
the data to the model. Strong broad Fe-K and Fe-L are again evident.

It is plausible that the photon index can change between observations
but not that the inclination can vary significantly, for such an inner
region immediately around the black hole. Examination of the spectrum
shows that the previous inclination of around 58 deg is the best fit
for the new data above 1.5~keV. The higher inclination is driven by a
slight excess around 1~keV. As seen in the lower panel of Fig.~3, the
reflection is complex in this region. Inclusion of another reflection
component with the same parameters as the first except for a higher
ionization parameter $\xi=990^{+1340}_{-140}\ergcmps$ provides a better fit
($\chi^2=308/337$; Fig.~5). It yields an inclination of
$57.6\pm 2.5$ deg, very close to that indicated by the 2008
results. We interpret this to mean that the surface of the disc has
regions of different density. (The best fits to the 2008 data also
required a similar additional high ionization component, Fabian et al 2009,
Zoghbi et al 2010.)

The above fits use the relativistic-blurring kernel {\sc kdblur} for
strightforward comparison to earlier work. The inner radius measured
indicates the extent of gravitational redshift required by the
spectrum.  The relativistic-blurring code {\sc Kerrconv} enables us to
measure the black hole spin, for which we obtain $a>0.997$, similar
to that found by Fabian et al (2009) and Zoghbi et al (2010).  The
emissivity index of the reflection then has $q\sim 6.4$. The
assumption made in determining the spin in this way is that the inner
radius, $r_{\rm in}$ from which reflection is detected corresponds to
the innermost stable circular orbit (ISCO) around the black hole. The
low value of $\xi$ found for the dominant reflection component means
that the disc within $2r_{\rm g}$ is dense, and thus spatially
thin. Under those conditions we expect that $r_{\rm in}$ is close
to the ISCO (Reynolds \& Fabian 2008). Matter falling inside the ISCO
on plunge orbits quickly drops to a much lower density where it
becomes completely ionized, so producing no iron reflection features.

\subsection{Timing and Variability}

A lightcurve of the whole observation is shown in Fig.~6 in several
energy bands, with higher resolution (200\,s bins) of the whole band
in Fig.~7. The source continues to show rapid variability with several
sharp spikes of emission. We see several instances where the emission
rises and drops by a significant amount (factor of two) on a timescale
of just a few 100\,s.

The variability is more pronounced at lower energies. This is shown in
Fig.~8, where the root-mean-square fractional variability is plotted
and compared with that of earlier observations.  The count rate is too
low to makes separate fits to the variations in 2011. The drop in
fractional variability around 0.9~keV evident in Fig.~8 has two
possible interpretations. 

Firstly, the flux could be low enough for a thermal component in the
host galaxy to be relevant. In this case (narrow) Fe-L line emission
would be significant at this energy and such a spatially-extended
component would not vary. A thermal component is not, however,
required by spectral modelling. (The count rate is too low for any
meaningful RGS analysis.) Secondly, the two component reflector model
allows for spectral variability if the components vary on different
timescales. Since the higher ionization component dominates over the
0.7--1.2~keV band, it seems that this is the less variable one. We
have therefore investigated a model in which the ionization parameter
of the lowly ionization component varies up to $\xi=15\ergcmps$. The
fractional change is shown in Fig.~9. The model is not a fit. We do
not pursue this further as more parameters could vary (both values of
$\xi$, normalisations, $\Gamma$ etc) in possibly complex and related
ways, but the trend is remarkably similar to the observed values.

\section{Light Bending Interpretation}

Our results show that most of the reflection originates from within a
radius of $2r_{\rm g}$ where strong gravitational light bending must
occur, enhancing the strength of the reflection component relative to
the direct continuum (Martocchia \& Matt 1996; Miniutti et al
2004). We now examine what this means for the location of the primary source.

The iron line emissivity profile of the accretion disc during the
observation was obtained by fitting the photon counts (normalisations)
of relativistically-broadened emission lines (from the single {\sc
  reflionx} X-ray reflection model convolved with the {\sc kdblur}
relativistic blurring kernel) originating from successive annuli in
the accretion disc to the observed iron K line in the reflection
spectrum (Wilkins and Fabian 2011). The emissivity profile is
approximated by a power law with a steep index of around 8
(Fig.~10). Further weak emission beyond $5 r_{\rm g}$ with an index of
between 3 and 4 over the outer regions of the disc is possible but is
poorly constrained. The results are consistent with fitting a single,
broadened, emission line profile ({\sc laor}) with a power-law
emissivity profile over the entire disc. The steep fall-off in
emissivity over the inner part of disc results in the flux emitted
from the outer disc regions being low, so the emissivity profile is
less well constrained here.  The emissivity profile has changed
substantially from that found in XMM-Newton EPIC pn spectra obtained
in 2008 January (Wilkins and Fabian 2011), where fitting reflection
from successive annuli in the accretion disc suggested a twice-broken
power law form for the emissivity profile, with a steep index of 7.8
out to a radius of $5 r_{\rm g}$, where the profile flattened to an
index of zero out to around $30 r_{\rm g}$ before tending to an index
of 3 over the outer regions of the disc.  Now almost 80 per cent of
the emission orginates from within $2 r_{\rm g}$, compared with 50 per
cent from that region in 2008 (see Fig.~10). The conclusions are
little changed if the emissivity index for the double reflector model
($q=6.5$) is adopted.

\begin{figure}
  \centering
  \includegraphics[width=1.1\columnwidth,angle=-0]{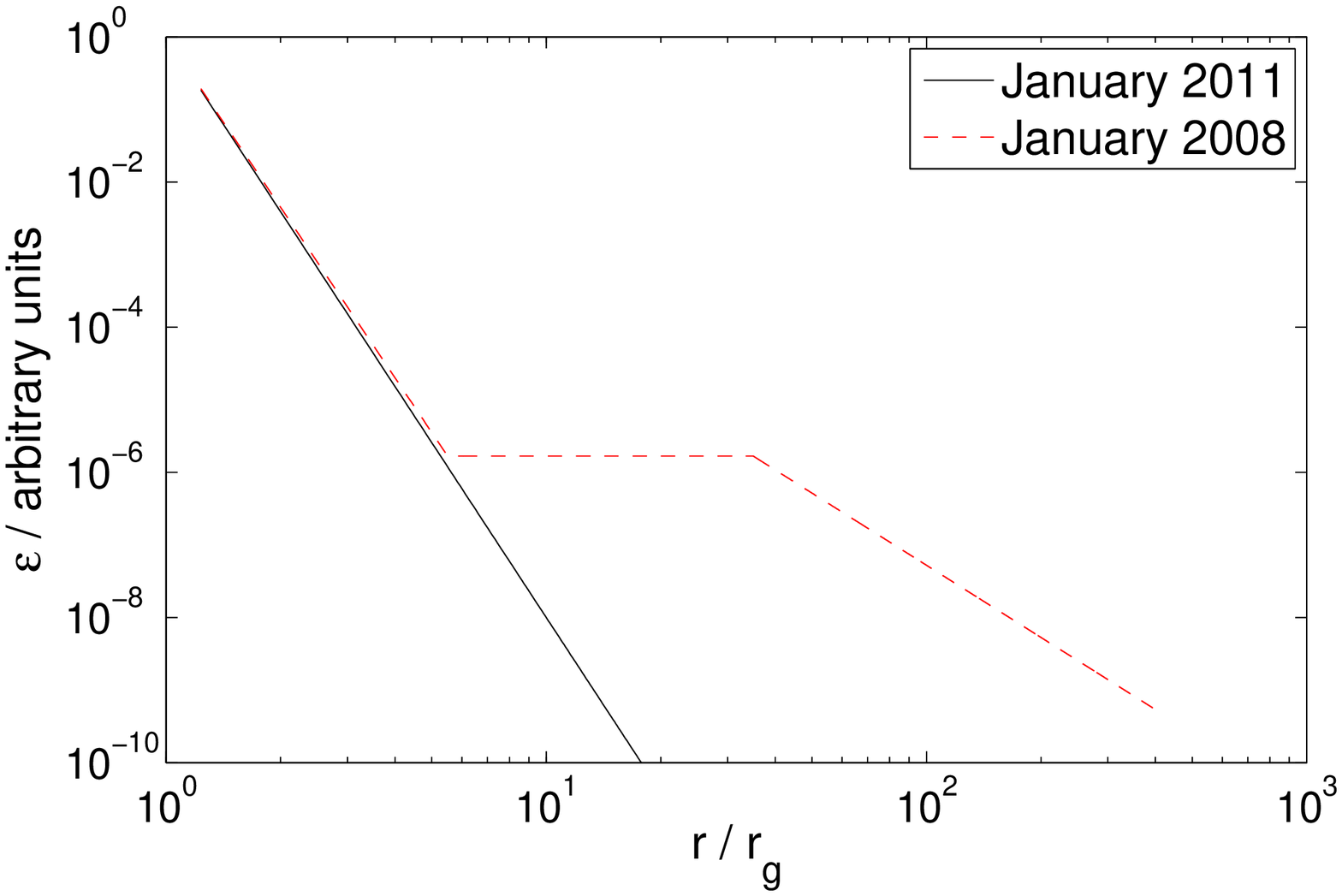}
  \includegraphics[width=1.\columnwidth,angle=-0]{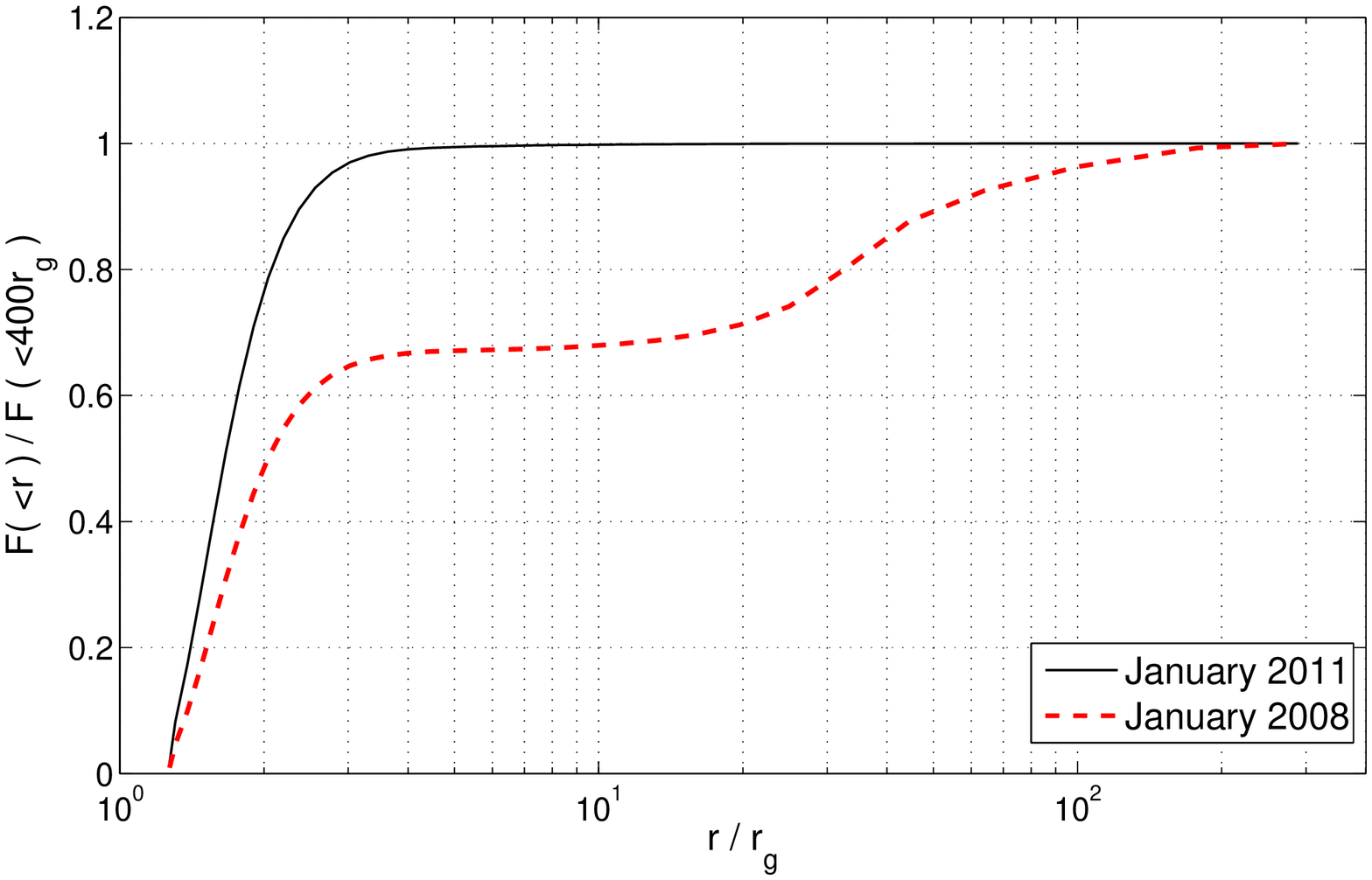}
  \caption{Top: Emissivity vs radius for the single reflector models
    fitting the spectra from 2011 (black) and 2008 (dashed
    red). Bottom: Cumulative flux observed at infinity, as a function
    of disk radius, due to emissivity profiles shown in the top
    figure.}
\end{figure}
\begin{figure}
  \centering
  \includegraphics[width=1.\columnwidth,angle=-0]{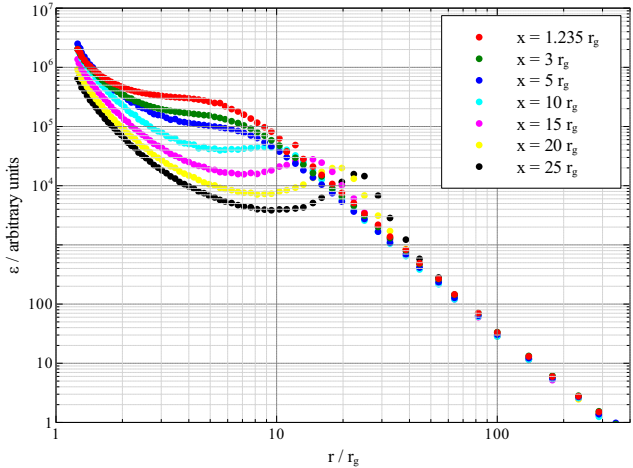}
  \caption{Theoretical emissivity profiles for point sources
    (equivalent to ring sources due to the axisymmetry of the
    spacetime) at a height of $5r_{\rm g}$ with varying radius corotating with
    the disc below. In the context of these profiles, the twice-broken
    power law emissivity profile obtained from the January 2008
    observations can be understood in terms of an extended source a
    low height above the disc which is  formed from the sum of
    such profiles.  }
\end{figure}
\begin{figure}
  \centering
  \includegraphics[width=1.\columnwidth,angle=-0]{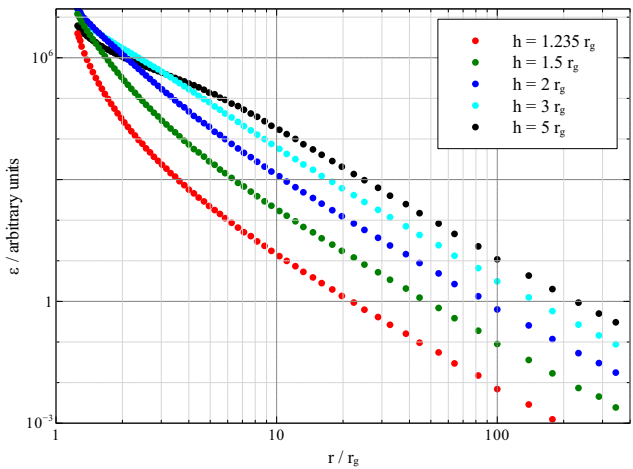}
  \caption{Theoretical emissivity profiles due to point sources at
    varying height on the rotation axis above the disc plane,
    suggesting that the once-broken power law emissivity profile with
    inner index 8 or more, quickly changing to $\sim 6$ and then to an
    index of between 3 and 4 at a radius of $5r_{\rm g}$ is produced
    by a primary source confined close to the rotation axis at a
    height less than $1.5r_{\rm g}$.  }
\end{figure}
\begin{figure}
  \centering
  \includegraphics[width=0.9\columnwidth,angle=-0]{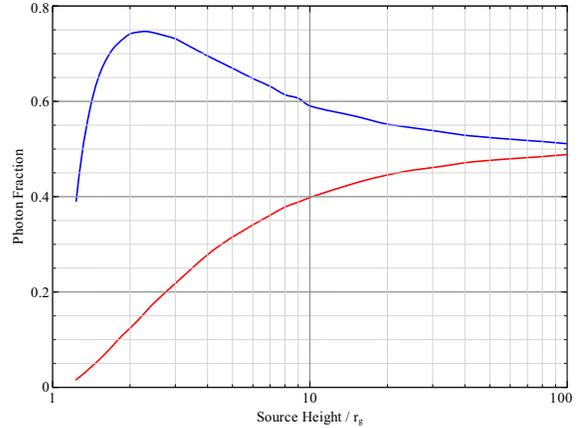}
  \caption{Fraction of photons escaping outward (red lower curve) or
    inward striking the disk (blue upper), determined from ray-tracing
    simulations.}
\end{figure}

The emissivity profiles are compared to theoretical predictions
obtained from ray tracing simulations. Rays are traced from an
isotropic point source at height $h$ above the plane of the accretion
disc, either at rest on the rotation axis or offset from the axis,
co-rotating with the disc below. Rays are traced from the source in
the Kerr spacetime around the central black hole, until they reach the
accretion disc in the equatorial plane, where their positions and
redshifts are recorded. The emissivity profile is found by counting
the total (redshifted) energy of rays landing in each bin, which is
proportional to the square of the redshift as both the energy of
individual photons and photon arrival rates are affected (the
emissivity profile is defined as the power emitted from the disc per
unit area). The energy is divided by the area taking into account
special and general relativistic effects due to the disc orbiting
close to the black hole.

From the emissivity profile obtained from the 2008 January  observation
along with constraints from studying the X-ray variability and
reverberation (Zoghbi et al 2010), it would appear that the accretion
disc is illuminated by an X-ray source close to the disc, while
extending radially to around $20 r_{\rm g}$ (Fig~11). Classically, the
emissivity profile is expected to be flat in the region below the
source where $r \ll h$, while tending to $r^{-3}$ when $r \gg h$, where
the flux received by the disc from the source falls off as the inverse
square of the distance with a further factor of r arising from the
cosine of the angle projecting the ray normal to the disc plane. The
reflected flux is enhanced over the inner region of the disc by
relativistic effects where rays are focussed towards the central black
hole while photons are blueshifted travelling inwards and the
spacetime in the disc is warped, increasing the disc area, steepening
the emissivity profile here.

The current observation, however, reveals an emissivity profile
explained by a compact source, confined to a small region around the
rotation axis and close to the black hole. The source is now required
to be at a height less than $1.5 r_{\rm g}$ above the disc plane to
explain the observed steepening over the inner region (Fig.~12). With
the source this close to the black hole, there is no flattened region
in the emissivity profile. Rather, the relativistic effects close to
the black hole steepen the inner region of the profile out as far as
the region where $r \gg h$, so the emissivity profile takes the form of
a once-broken power law from the steep inner profile to a profile
slightly steeper than $r^{-3}$ (with the slight steeping from the
classical case due to rays being focussed towards the black hole) over
the outer regions of the disc.  Ray tracing simulations also
illustrate the variation in the reflected flux observed compared to
that in the power law continuum as a function of the source height, by
counting the photons in the simulation that hit the accretion disc
compared to those which are able to escape to infinity (Fig.~13). A
similar result is obtained when considering the variation in reflected
and continuum flux as a function of source radius from the rotation
axis. The substantial drop in the observed power law continuum between
the 2008 January observation and that of 2011 supports the hypothesis
that the primary X-ray source has collapsed down to a region close to
the black hole such that relativistic effects focus the majority of
the rays towards the black hole and on to the disc, so a greater
fraction of the emitted X-rays are reflected from the disc, while very
few are able to escape as part of the continuum. The source thus
becomes reflection dominated.

\section{Alternative models}

\begin{figure}
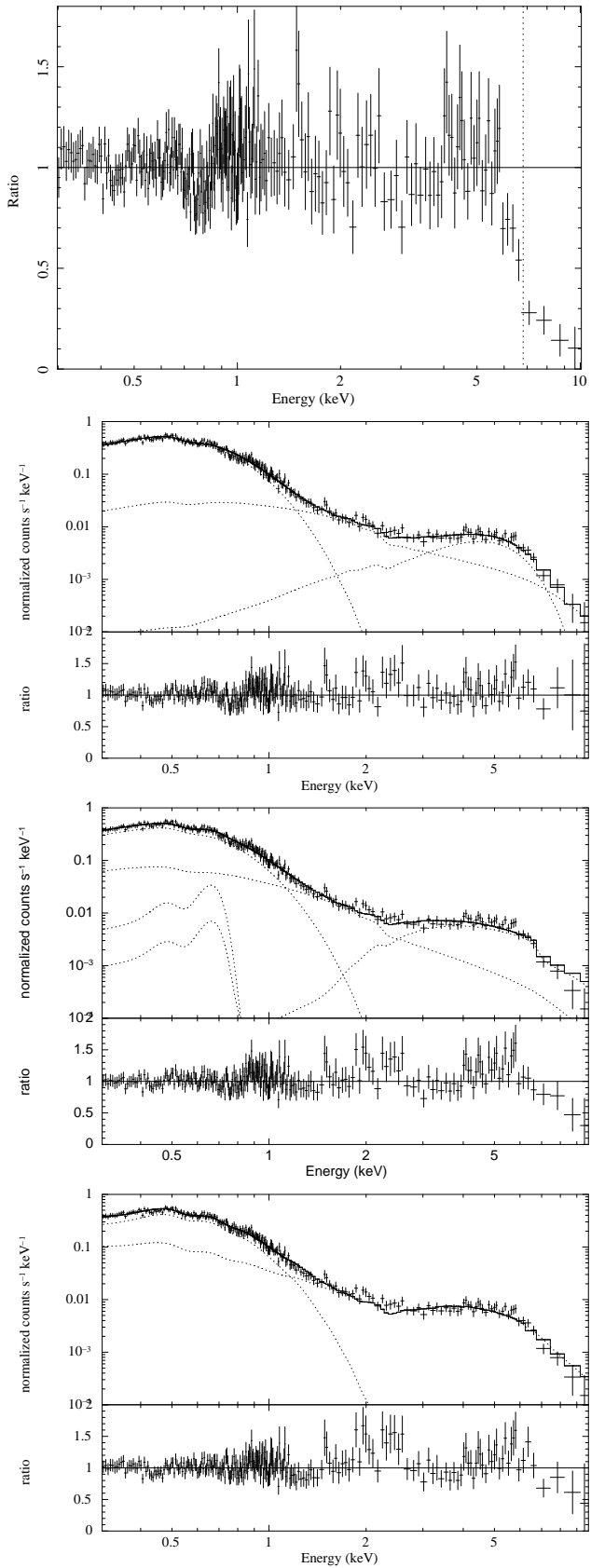

  \centering
  \includegraphics[width=0.68\columnwidth,angle=-90]{bbpo.ps}
  \includegraphics[width=0.65\columnwidth,angle=-90]{bbpogaadd.ps}
  \includegraphics[width=0.65\columnwidth,angle=-90]{pcvfe.ps}
  \includegraphics[width=0.65\columnwidth,angle=-90]{zxipcfadd.ps}
  \caption{Top: Ratio of XMM 2011 spectrum to a simple blackbody plus
    powerlaw model fitted over the band 0.3--5~keV. Upper Centre: Blackbody
    plus powerlaw and gaussian model. Lower Centre: Partial-covering model;
    the absorbing column density is $8\times 10^{21}\psqcm$ and Fe
    abundance is 100. Bottom: Partially-ionized absorber model. Note the
  poor fit above 5~keV. }
\end{figure}

The spectrum of the source is shown in Fig.~14 as a ratio to the
power-law plus blackbody model which best fits over the energy range
0.3--6~keV (i.e. omitting the drop in the spectrum). The blackbody has
a temperature of $0.123\pm0.002\keV$ and the powerlaw index is
$\Gamma=0.35\pm0.04$. The spectral drop beginning at 6 keV is
clear. The rest frame energy of the neutral iron absorption edge at
7.1~keV is indicated by a dotted line in the figure and is above the
energy of the observed drop. The spectral index of the power-law
component, which dominates from 1.5--6~keV, is flatter than any
plausible Comptonization process and so is not a reasonable solution. 

The emission from 1.5--10~keV is well fit by a gaussian component with
a central energy of $4.70^{+0.24}_{-0.34}\keV$ and dispersion of
$1.52^{+0.32}_{-0.23}\keV$ (Fig.~14, upper centre).  Next, a
partial-covering solution is shown in Fig.~14 (lower centre) with
$\Gamma=2.33\pm0.06$. It requires 92 per cent covering by neutral gas
of column density $N_{\rm H}=8\times 10-^{21}\pcmsq$ with an iron
abundance exceeding 80 times the Solar value (90 per cent confidence
level).

Both of these two models also require a blackbody component with a
temperature of $0.116\pm0.003\keV$ to account for the flux below 1~keV
($\chi^2/{\rm dof}=316/345$ and $336/345$ in the 0.3--10~keV band for
the Gaussian and partial covering models, respectively). This ``soft
excess'' component is common in AGN and is unlikely to be a true
blackbody associated, say, with the accretion disc since it shows a
similar temperature over a wide range of object independent of black
hole mass or accretion rate (Gierlinski \& Done 2004; Crummy et al
2006), contrary to the expectation of accretion theory. The simplest
interpretation of such a (roughly) constant energy component is that
it is due to atomic processes, in either absorption or emission. The
lack of any sharp defining features requires that there is significant
relativistic blurring which brings us back to strong
reflection.

The strength of the drop at 6--7~keV, which requires a very extreme
iron abundance (much higher than required to explain the earlier spectra,
Boller et al 2002), makes any partial-covering model implausible, in
our view. This model was also the worst fit, particularly in the
4--6~keV band. 

A broad gaussian component could be due to a down-scattered iron
emission line, say from FeXXV at 6.7~keV. To be down-scattered to
4.7~keV requires about 40 Compton collisions (a Thomson depth of about
6), which would give a dispersion of about 0.4 keV. The fitted
dispersion of 1.5~keV could then be explained by scatterer which is at
$kT\sim 0.4\keV$. Such a medium would have to be completely ionized in
order that it does not absorb the emission. This does not however
explain the source of such an intense iron emission line in the first
place, nor will it easily fit the previous observations.

Finally, a model with a power-law, $\Gamma=2.87$, and blackbody
components and two ionized absorbers ({\sc Xspec} model{\sc zxipcf}
$N_{\rm H,1}=2.6\times 10^{23}\psqcm$, $\log \xi=3.56$, $f=1$; $N_{\rm
  H,2}=2.2\times 10^{23}\psqcm$, $\log \xi=1.79$, $f=0.975$, where $f$
is the covering fraction) can give a fair fit ($\chi^2=339$ for 343
d.o.f; Fig.~14, lower panel). The model has Fe-L and other absorption
lines.  No such lines or evidence for an ionized absorber were seen in
the deep RGS spectrum from 2008 when the source was brighter (Blustin
\& Fabian 2009). 

No absorber model for the bright state spectrum of 1H\,0707-495 has
yet accounted for the structure around both 7~keV and 1~keV in any
self-consistent manner, such as is achieved with the blurred
reflection model. Gallo et al (2004), for example, added a gaussian
emission line at 0.92~keV. The spectrum we report on here appears to
be just a more redshifted, and fainter, version of the bright state
spectrum (Fig.~2). It does not make sense to us that the broad
features in the low-state spectrum have a different physical origin
to those in the high state.

\subsection{An extended scattering region?}

A major discovery from the 2008 XMM-Newton data was a $\sim 30\s$ soft
lag found at high frequencies by Fabian et al (2009) and Zoghbi et al
(2010).  The lag was interpreted as the reverberation time delay due
to the response of the reflection component, which dominated the soft
flux ($<1\keV$), to variations in the power-law component, which is seen
directly and dominates in the 1--3~keV band. Although the lack of any
power-law component and low soft count rate preclude any lag analysis
with the 2011 data, the interpretation is fully consistent with the
unobscured, inner reflection, model which we  use here to
understand the 2011 spectrum.

Miller et al (2010) have carried out a new timing analysis of the 2008
XMM-Newton data of 1H\,0707-495. They confirm the $\sim 30\s$ soft lag
but claim that the size of the lag at higher energies rules out the
inner reflection model and instead propose that the source is
surrounded by a $\sim 1000$~light sec radius scattering region. Holes
are envisaged in the scattering cloud such that direct flux does
emerge along our line of sight.  We do not replicate their result using
standard timing procedures, finding on the contrary that the spectral
dependence agrees well with our model (Zoghbi et al 2011). We are
unable to check their approach since it uses a new, but as yet
unpublished method.

Their model is motivated by the presence of a large positive low
frequency lag, in which the soft flux leads the harder flux. They
argue that this is due to time delays in a large scattering cloud
around the source. In their model, direct emission is seen 
dominating at soft energies below 1~keV while the scattered/reflected
flux dominates above that energy. No detailed model which fits the
spectral data is presented by Miller et al (2010), but we note that
their direct and scattered/reflected components dominate in the soft and
hard bands in the opposite way to those of the inner reflection model.
Fabian et al (2009) and Zoghbi et al (2010) comment that the low
frequency lags are similar to those commonly seen in Galactic binary
black hole systems, where they are not due to scattering within a
large cloud, and are likely to be due to accretion fluctuations
propagating inward in the disc and communicated to the central coronal
structure by magnetic fields from a range of radii.
 

In a model in which scattering occurs in a 1000~lt~s radius cloud,
then the power spectrum of the scattered/reflected component (hard
band) should be steeply attenuated above $4\times 10^{-4}\Hz$, due to
light travel time smearing in the cloud, relative to the power
spectrum of the direct component (i.e. the soft band). The observed
power spectra however show exactly the {\em opposite} behaviour (see
Fig.~5 of Zoghbi et al 2011). The soft bands have less high-frequency
power than the hard bands, which show no sign of attenuation due to
light travel time effects, indeed they even show an enhancement at
higher frequencies. Finally, we note that recent discoveries of soft
lags in other sources (Emmanoulopoulos et al 2011, Tripathi et al
2011, De Marco et al 2011, Zoghbi et al 2011) support the inner
reflection model for the 2008 data of 1H\,0707-495, and argue against any
special geometry or source inclination for such sources.

We infer that a model involving a large scattering cloud is an
untenable interpretation of the XMM-Newton data of 1H\,0707-495.  The
sharp changes in the light curve shown in Fig.~7 are strong evidence
against any significant scattering cloud being present in 2011.

\section{Summary and Discussion}

The low-state observation of 1H\,0707-495 studied here shows a
remarkable reflection-dominated spectrum and strong variability. We
obtain a consistent solution to the degree of relativistic blurring
required and lack of a power-law component if the black hole is
spinning very fast with spin parameter $a> 0.997$, and the irradiating
power-law source lies within $1r_{\rm g}$ of the black hole event
horizon, i.e. at a radius of $2r_{\rm g}$ from the singularity. To our
knowledge these are the first measurements {\em dominated} by emission from
so close to the black hole.

The low ionization parameter of the dominant reflection component
means that the disc remains dense right to the innermost radii. This
in turn implies that the disc remains thin. No strong poloidal
magnetic fields are therefore expected so there is unlikely to be any
strong Blandford-Znajek (1977) effect.

Orban de Xivry et al (2011) argue that the extreme properties of
Narrow Line Seyfert 1 galaxies as a class is due to the growth of
their black holes being dominated by internal secular evolution. This
leads to prolonged gas accretion and high black hole spin.

The main reduction in flux of 1H\,0707-495, which is of an order of
magnitude relative to 2008, occurs in the soft band, below 1~keV. The
hard band flux around 5~keV drops by only a factor of two. Part of the
soft flux drop is due to a change in the spectral index of the
irradiating continuum, from $\Gamma\sim3$ to $\Gamma\sim2.4-2.7$. The
energy density in soft disc photons incident onto the power-law source
will change as it approaches the black hole. Assuming that the
power-law emission is due to Comptonization from constant temperature
electrons, then $\Gamma$ reduces (the source hardens) if the soft
photon energy density reduces. It is plausible that the energy density
drops very close to the black hole, in support of our
interpretation. We shall examine this further with future detailed
calculations.

Several alternative interpretations of the low-state spectrum have
been studied. None offers a simple physical and consistent
interpretation.

\section*{Acknowledgements}
We thank the referee for a helpful report.  ACF thanks the Royal
Society for support. PU is supported by an STFC Advanced Fellowship
and funding from the European Community's Seventh Framework Programme
(FP7/2007-2013) under grant agreement number ITN 215212 ``Black Hole
Universe''. DG acknowledges support by NASA grant NNX09AN12G.


\begin{thebibliography}{}
\bibitem[]{} Ballo L., Giustini L., Schartel N., Cappi M.,
  Jim\'enez-Bail\'on E.,  Piconcelli E., Santos-Lle\'o, M., Vignali
  C., 2008, A\&A, 483, 137
\bibitem[]{} Blandford R.D., Znajek R.L. 1977, MNRAS, 179, 433
\bibitem[]{} Blustin A.J., Fabian A.C., 2009, MNRAS, 399, L169
\bibitem[]{} Boller T., Fabian A.C., Sunyaev R., Tr\"umper J., Vaughan
  S., Ballantyne D.R., Brandt W.N., Keil R., Iwasawa K., 2002, 329, 1
\bibitem[]{} Crummy J., Fabian A.C., Gallo L.C., Ross R.R., 2006,
  MNRAS, 365, 1067
\bibitem[]{} De Marco B., Ponti G., Uttley P., Cappi M., Dadina M.,
  Fabian A.C., Miniutti G., 2011, MNRAS submitted
\bibitem[]{} Emmanoulopoulos D., McHardy I.M., Papadakis I.E., 2011,
  MNRAS in press (arXiv:1106.606)
\bibitem[]{} Fabian A.C., Ballanytyne D.R., Merloni A., Vaughan S.,
  Iwasawa K., Boller T., 2002, 331, L35
\bibitem[]{} Fabian A.C., Miniutti G., Iwasawa K., Ross R.R., 2005,
  MNRAS, 361, 795
\bibitem[]{} Fabian A.C. et al., 2009, Nature, 459, 540 
\bibitem[]{} Fabian A.C., Miniutti G., Gallo L., Boller T., Tanaka Y.,
  Vaughan S., Ross R.R., 2004, MNRAS, 353, 1071
\bibitem[]{} Fabian A.C., et al 2009, Nature, 459, 540
\bibitem[]{} Gallo L.C., Tanaka Y., Boller T., Fabian A.C., Vaughan
  S., Brandt W.N., 2004, MNRAS, 353, 1064
\bibitem[]{} Gehrels N., 1986, ApJ, 303, 336
\bibitem[]{} Gierlinski M., Done C., 2004, MNRAS, 349, L7
\bibitem[]{} Grupe D., Komossa S., Gallo L.C., Fabian A.C., Larsson
  J., Pradhan A.K., Xu D., Miniutti G., 2008 ApJ, 681 982
\bibitem[]{} Martocchia A., Matt G., 1996, MNRAS, 282, L53
\bibitem[]{} Miller L., Turner R.J., Reeves J.N., Braito V., 2010,
  MNRAS, 408, 1928 
\bibitem[]{} Miniutti G., Fabian A.C., 2004, MNRAS, 349, 1435
\bibitem[]{} Orban de Xivry G., Davies R., Schartmann, M., Komossa S.,
  Marconi A., Hicks, E., Engel H., Tacconi L., 2011, MNRAS in press
  (arXiv:1104.5023)
\bibitem[]{} Kalberla P.M.W., Burton W. B., Hartmann D., Arnal E.M.,
  Bajaja E.,  Morras R., Pöppel W.G.L., 2005, A\&A, 440, 775
\bibitem[]{} Ponti G., Miniutti G., Cappi M., Maraschi L., Fabian
  A.C., Iwasawa K., 2006, MNRAS, 368, 903 
\bibitem[]{} Reynolds C.S., Fabian A.C., 2008, ApJ, 675, 1048
\bibitem[]{} Ross R.R., Fabian A.C., 2005, MNRAS
\bibitem[]{} Schartel N., Rodr\'iguez-Pascual P.M., Santos-Lle\'o, M.,
  Jim\'enez-Bail\'on E., Ballo L., Piconcelli E., 2010, A\&A 512 A75
\bibitem[]{} Tripathi S., Misra R., Dewangan G., Rastogi S., 2011, ApJL
  in press, (arXiv:1106.6i39)
\bibitem[]{} Vignali C., Piconcelli E., Bianchi S., Miniutti G., 2008,
  MNRAS, 388, 761
\bibitem[]{} Wilkins D.R., Fabian A.C., 2010, MNRAS, 414, 1269 
\bibitem[]{} Zoghbi A., Fabian A.C., Uttley P., Miniutti G., Gallo
  L.C., Reynolds C.S., Miller J.M., Ponti G., 2010, MNRAS, 401, 2419
\bibitem[]{} Zoghbi A., Uttley P., Fabian A.C., 2011, MNRAS, 412, 59
\bibitem[]{} Zoghbi A., Fabian A.C., 2011, MNRAS submitted
\end{thebibliography}

\end{document}